\documentclass{emulateapj}

\usepackage{graphicx}
    \usepackage{epsfig}
    \usepackage{natbib}
    \usepackage{xcolor}
    \usepackage{float} 
    \usepackage{amsmath}
    \usepackage{pxapj}

\shorttitle{Mrk1239} \shortauthors{Pan et al.}

\slugcomment{}

\begin{document}


\title{Mrk 1239: a Type-2 Counterpart of Narrow-line Seyfert-1?}


\author{Xiang\ Pan\altaffilmark{1},  Hongyan Zhou\altaffilmark{1,2}, Chenwei Yang\altaffilmark{1}, Luming Sun\altaffilmark{3}, Paul S. Smith\altaffilmark{4}, Tuo Ji\altaffilmark{1}, Ning Jiang\altaffilmark{2}, Peng Jiang\altaffilmark{1}, Wenjuan
Liu\altaffilmark{5}, Honglin Lu\altaffilmark{2}, Xiheng Shi\altaffilmark{1}, Xuejie, Dai\altaffilmark{1}, Shaohua Zhang\altaffilmark{6} } 
\altaffiltext{1}{Key Laboratory for Polar Science, MNR, Polar Research Institute of China, 
451 Jinqiao Road, Shanghai, 200136, China; $^\ddagger$Correspondence:
zhouhongyan@pric.org.cn, yangchenwei@pric.org.cn}
\altaffiltext{2}{Key Laboratory for
Researches in Galaxies and Cosmology, Department of Astronomy,
University of Sciences and Technology of China, Chinese Academy of
Sciences, Hefei, Anhui, 230026, China}  
\altaffiltext{3}{School of Physics and Electronic Information, Anhui Normal University, 241002, Wuhu, Anhui, People's Republic of China}
\altaffiltext{4}{Steward Observatory, The University of Arizona, Tucson, AZ 85721, USA}
\altaffiltext{5}{Yunnan Observatories,
Chinese Academy of Sciences, Kunming, Yunnan 650011, China }
\altaffiltext{6}{Shanghai Key Lab for Astrophysics, Shanghai Normal University, 200234, Shanghai, People's Republic of China}

\begin{abstract}
We present new spectrophotometric and spectropolarimetric observations of Mrk 1239, one of the 8 prototypes that defines type-1 narrow-line Seyfert galaxies (NLS1s). 
Unlike the other typical NLS1s though, a high degree of polarization ($P\sim$5.6\%) and red optical-IR ($g-W_4$ = 12.35) colors suggest that Mrk 1239 is more similar to type-2 active galactic nuclei like NGC 1068. Detailed analysis of spectral energy distribution in the UV-optical-IR yields two components from the nucleus: a direct and transmitted component that is heavily obscured ($\ebv \approx 1.6$), and another indirect and scattered one with mild extinction (\ebv $\sim$ 0.5). Such a two-ligh-paths scenario is also found in previous reports based on the X-ray data. Comparison of emission lines and the detection of \hei*$\lambda$10830 BAL at [-3000,-1000] \kms indicates that the obscuring clouds are at physical scale between the sublimation radius and that of the narrow emission line regions. The potential existence of powerful outflows is found as both the obscurer and scatterer are outflowing. Similar to many other type-2s, jet-like structure in the radio band is found in Mrk 1239, perpendicular to the polarization angle, suggesting polar scattering. We argue that Mrk 1239 is very probably a type-2 counterpart of NLS1s. The identification of 1 out of 8 prototype NLS1s as a type-2 counterpart implies that there can be a substantial amount of analogs of Mrk 1239 misidentified as type-1s in the optical band. Properties of these misidentified objects are going to be explored in our future works.
\end{abstract}



\section{Introduction}
Based on whether or not we detected featureless continuum and broad emission lines, active galactic nuclei (AGNs) are categorized into type-1s and type-2s respectively. The identification of scattered broad emission lines in type-2 AGNs then led to an orientation-based model (Antonucci 1993) that unifies these two types of AGNs. In this model, type-1 and type-2 AGNs are intrinsically the same, with a similar central engine (continuum-emitting disks and broad emission line regions) in the center, surrounded by dusty tori at the equatorial plane. Objects observed face-on are type-1s, and edge-on objects in which the central engines are blocked are called type-2s.

Such a unification scheme remains to be one of the greatest successes in our understanding of the AGN phenomena to date. However, follow-up works found evidence that there can be intrinsic differences between type-1 and type-2 AGNs. 
After correcting selection bias, Ricci et al. (2015) found that the fraction of obscured Compton-thin (log $N_{\rm H}$ = 22-24) AGNs decreases with increasing X-ray luminosity, from  46$\pm$3\% (for log$L_{14-195\ {\rm keV}}$ = 40--43.7) to 39$\pm$3\% (for log$L_{14-195\ {\rm keV}}$ = 43.7--46), indicating that the intrinsic luminosity or the distribution of obscuring contents in type-1 and type-2 AGNs can possibly be different. For AGNs at higher accretion rates, strong radiative feedback tends to push obscuring materials further away, leading to a decrease of covering factors of obscurers discovered at higher accretion rates (Ricci et al. 2017a).
It is also proposed that the surrounding dusty tori can be clumpy, meaning that there are type-1 AGNs even if observed edge-on (Nenkova et al. 2008), and their covering fractions in type-2 AGNs tend to be larger than in type-1s (Elitzur 2012). 
On the other hand, the fractions of obscured AGNs are found to be larger at higher $z$ (Hasinger 2008; Buchner et al. 2015), indicating the over-abundance of obscuring contents in AGNs at higher redshifts (Treister \& Urry2006).
By studying the X-ray emission of galaxy mergers, often thought to activate the rapid accretion on to supermassive black holes (the start of the AGN stage), Ricci et al. (2017b) found that the fraction of compton-thick object is much higher (65$^{+12}_{-13}$\%) than for local hard X-ray selected AGNs (27$\pm$4\%). Thus there can be potential evolution from the type-2 population to the type-1s.

A narrow-line Seyfert-1 galaxy (NLS1) is a subclass of AGNs having narrower broad emission lines at full width half maximum ($FWHM$) $ \lesssim 2000$ km s$^{-1}$, strong \feii multiplets, steep and highly variable X-ray emission, among other properties (Leighly 1999a, 1999b; Zhou et al. 2006). NLS1s have smaller black hole mass and extremely high accretion rates as compared to normal Seyferts, and thus are considered to be early black holes in their quick growth stage (e.g., Mathur 2000; Grupe \&
Mathur 2004; Zhou et al. 2006; Komossa et al. 2008; Xu et al.2012). So they are important in understanding both the unification and evolution of AGNs.
Having continuum and broad emission lines obscured, type-2 AGNs are more difficult to detect than type-1s under the same luminosities and distances. This kind of selection effect makes the sample sizes of type-2s significantly smaller than that of type-1s, especially in the optical band.
In terms of type-2 counterparts of narrow-line Seyfert galaxies (NLS2), an additional difficulty is that the broad emission lines of NLS2s are heavily obscured, making the identification through conventional criteria ($FWHM_{\rm BEL} \lesssim 2000$ \kms) hard to achieve (Dewangan \& Griffiths 2005).

Recently, a NLS2 SDSS J120300.19+162443.7 (J1203+1624 hereafter) is identified via optical-IR spectrophotometry and spectropolarimetry. A high degree of polarization is found in the optical band, along with significant polarized broad emission lines, very similar to the prototype of Seyfert-2 galaxies, i.e. NGC 1068. Spectral analysis found out that the scattered featureless continuum and broad emission lines in J1203+1624 are so strong that they can be directly measured in the optical spectrum. Such a discovery suggests that the obscured nuclei in some NLS2s can be uncovered in scattered light (Pan et al. 2019). Based on properties of J1203+1624, we then start searching for other NLS2s. Interestingly, we find out that Mrk 1239, one of the 8 prototypes that define NLS1s (Osterbrock \& Pogge 1985), is very likely one. Archival data are collected, and follow-up observations are carried out, which are described in \S2. As compared in \S3, the high degree of polarization in Mrk 1239 is similar to typical type-2s rather than NLS1s. The red optical-IR color of Mrk 1239 strengthens such resemblances (\S4). And the UV-optical-IR spectral energy distribution (SED) is then decomposed in \S5, yielding an obscured and a scattered light path, similar to the previous findings based on the X-ray data. We then study the emission and absorption line spectrum in \S6, in which properties of the obscuring and scattering clouds are inferred. In conclusion, we discuss the classification of Mrk 1239 and implications in \S6 and \S7 respectively. Throughout the paper, all uncertainties are at 1-$\sigma$ level if not stated specifically. We adopt the standard $\Lambda$CDM model with $H_0 = 70$~\kms Mpc$^{-1}$, $\Omega_{\rm M}=0.3$, and $\Omega_{\Lambda}=0.7$.

\section{Observations, Data Reduction, and Data Collection.}
There are lots of historical observations of Mrk 1239, a bright nearby emission line galaxy.
A high-resolution optical image of Mrk 1239 was obtained by $Hubble~Space~Telescope$($HST$)\footnote{Based on observations made with the NASA/ESA $HST$, obtained at the Space Telescope Science Institute, which is operated
by the Association of Universities for research in Astronomy, Inc.,
under NASA contract NAS 5-26555.}/WFPC2 on 1995 February 27 in the F606W band (at $\sim$ 0.6\micron, Figure 4). We then collected images and fluxes from several surveys in the UV-optical-IR, including $Galaxy Evolution Explorer$ (GALEX, Morrissey et al. 2007), Sloan Digital Sky Survey (SDSS, York et al. 2000), UKIRT Infrared Deep Sky Survey (UKIDSS, Lawrence et al. 2007), Two-Micron All-Sky Survey Extended Source Catalog(2MASS XSC, Jarrett et al. 2000), and $Wide-field Infrared Survey Explorer$ ($WISE$, Wright et  al. 2010). {Photometric data of typical AGNs shown in this paper are also obtained in these surveys.}

Four $IUE$ spectra of Mrk 1239 were taken on 1987 November 10, 1988 April 16, 1988 May 25 and 1988 May 28 respectively (Crenshaw et al. 1991), with a large oval aperture (10\arcsec~ by 20 \arcsec),
 spanning a wavelength range of 1150-1979\AA. The combined $IUE$ spectrum is at a flux level similar to the $GALEX$ far-UV photometry (Figure 5).
An archival optical spectrum of Mrk 1239 is released in the 6dF Galaxy Survey (Jones et al. 2009), which is observed on 2004 April 26 with a large aperture of 6\arcsec.7 and covering a wavelength range of 4000-7600\AA. 
A near-IR (NIR) spectrum (0.8-2.4 $\mu$m) of Mrk 1239 is also available, which was observed by IRTF/SpeX on 2002 April 21 and 2002 April 23 (Riffel et al. 2006).
We performed optical spectropolarimetry of Mrk 1239 with MMT/SPOL (Schmidt et al. 1989) on 2016-Apr-4 and 2016-Apr-5 to study the scattered AGN radiation. A total exposure time of 920s is taken, with a 600 g mm$^{-1}$ grating. At a slit width of 3\arcsec, the SPOL data cover a wavelength range of [4200,8200]\AA~at a resolution of $\sim$7\AA. Since both the 6dF and the MMT/SPOL data were obtained with relatively large apertures, we calibrated these two spectra with low-order polynomials to match the optical fluxes based on SDSS petrosian magnitudes.
We also observed Mrk 1239 on 2013 February 23 in the NIR with the TripleSpec spectrograph on the Palomar 200-inch Hale telescope. Four 180s-exposures were teken, in an A-B-B-A dithering mode, at a seeing of 1\arcsec.1. A nearby A0V star is then observed for flux calibration and correction of telluric absorption. 1D spectrum with a spectral resolution of $\sim 3000$ was then extracted using the XSPEX tool package (Cushing et al. 2004). The TripleSpec data generally agree with the archival Spex data. The NIR absorption and emission lines are better resolved in the TripleSpec data for its higher spectral resolution. Mid-infrared (MIR) spectrascopy of Mrk 1239 is performed with the Infrared Spectrograph (IRS) on the $Spitzer Space Telescope$ on 2006 December 19. We obtained the combined IRS data from the “Cornell Atlas of Spitzer/IRS Sources” (CASSIS; Lebouteiller et al. 2011).
By checking multi-epoch observations of Mrk 1239 in the UV ($IUE$ spectra), optical (Catalina Sky Survey
\footnote{The standar deviation of the 362 epochs (2005 April 4 -- 2013 May 29)
 Catalina (ID: CSS\_J095219.1-013643) V-band photometry is 0.069 mag, consistent with measurement uncertainty of 0.050 mag, indicating negligible variability.}, Drake et al. 2009) and in the IR ($WISE$), variability was found to be negligible.
In addition, Mrk 1239 was observed by Keck on 1999 October 25, with the Echelette Spectrograph and Imager (ESI). A 0\arcsec.75-wide slit was applied, covering a wavelength range of 4000-11000\AA~ at a resolution of $\sim8000$. We reduced the raw data with XIDL and extracted spectrum of the nucleus with a relatively small aperture of 0\arcsec.72 to study the emission and absorption lines. We calculated the fluxes within the 0\arcsec.75$\times$0\arcsec.72 aperture based on $HST$ and SDSS images in the optical band, and a low-order polynomial is then applied to calibrate the flux of the Keck/ESI data.

\section{High Degree of Polarization}
\begin{figure}
\includegraphics[width=0.49\textwidth]{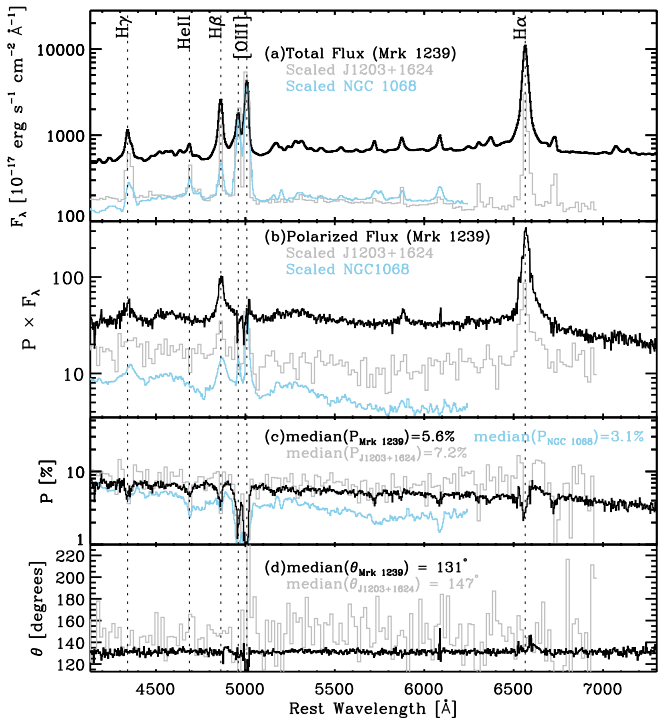}
\caption{
The total flux spectra ($F_\lambda$) of Mrk 1239 are shown in panel (a) as black lines, in comparison with J1203+1624 (grey line, Pan et al. 2019) and NGC 1068 (cyan line, digitized from Figure 2 of Antonucci 1993). 
Polarized flux spectra of the three targets are shown in panel (b).
In panel (c), degree of polarization ($P$) is shown as a function
of wavelength, with median values in the common wavelength range of [4000-6000]\AA~for these objects printed in the panel. The polarization angle ($\theta$) is displayed in panel (d), which is around 131$^\circ$ for both the continuum and the broad emission lines (BELs) of Mrk 1239.}
\end{figure}

\begin{figure}
\includegraphics[width=0.49\textwidth]{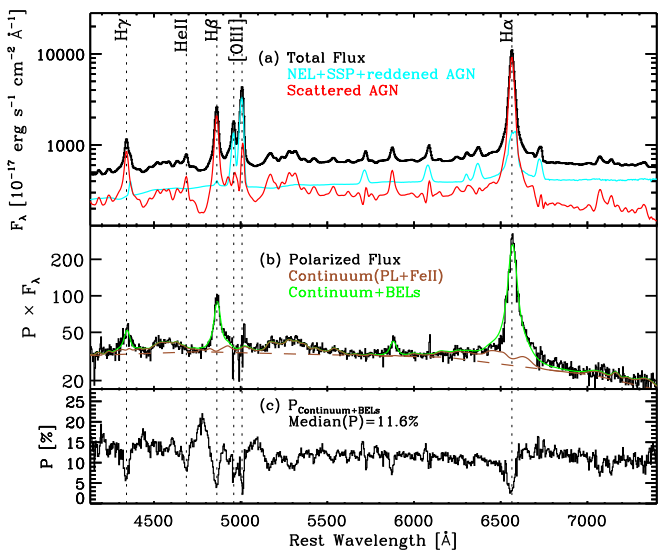}
\caption{
The total flux spectra ($F_\lambda$) of Mrk 1239 is decomposed into scattered AGN radiation (red line) and nonscattered component (starlight, transmitted AGN radiation, narrow emission lines) in panel (a). 
Polarized flux spectra are shown in panel (b), in which the scattered continuum (broken power law and \feii multiplets) and emission line models are shown in brown and green lines respectively.
In panel (c), the intrinsic degree of polarization ($P$) is shown as a function of wavelength, with a median value of 11.6\%.}
\end{figure}

The degrees of polarization in Seyfert-1 galaxies are generally equal or below 0.5\% (Martin et al. 1983). Despite the original identification of Mrk 1239 as a type-1 object, though, a high degree of continuum polarization of $P\sim3.35\%$ was discovered in it (Goodrich et al. 1989). High $P$ value of Mrk 1239  was also found in our spectropolarimetric data. As shown in Figure 1, continuum and broad emission lines can be clearly seen in the polarized flux (panel b) of Mrk 1239, the prototype Seyfert-2 NGC 1068 and the NLS2 J1203+1624. In the high-quality polarized flux of Mrk 1239 and NGC 1068, \feii multiplets are also detected. The observed $P$ in a common wavelength range of [4000-6000]\AA~ for all the three targets are also at a similar level of $\sim 5\%$(Figure1(c)).  We then follow previous works and remove nonscattered components of Mrk 1239 (starlight, transmitted AGN radiation, and narrow emission lines) from the total flux spectrum (see \S5 and \S A2 for a detailed description). The intrinsic $P$ of Mrk 1239 is then $P>10\%$ in the optical band (Figure 2(c)), increasing with decreasing wavelength. A high intrinsic polarization value similar to typical type-2s detected in a prototype NLS1 was quite unexpected. In fact, among the 8 prototype NLS1s in Osterbrock \& Pogge (1985), Mrk 1239 shows the highest level of continuum  polarization, and 6/8 of the targets are polarized at levels below 1\%.

\section{Red Optical-IR Colors}

The high degree of polarization detected in Mrk 1239 is unique among typical NLS1s; however, Mrk 1239 is quite similar to the prototype Seyfert-2 NGC 1068 and the NLS2 J1203+1624. Color-color diagram is a powerful tool in distinguishing type-1 and type-2 AGNs; we thus compared Mrk 1239 with some typical objects and samples in it. The synthetic rest-frame color $g-W_4$ is adopted, as an indicator of the relative flux between the optical $g$ and MIR $WISE\ W_4$. And the synthetic $g-J$ color indicates the relative flux between the optical $g$ band and $J$ band in the NIR. Besides NLS1s from  Zhou et al. (2006), type-2s, including the prototype Seyfert-2 NGC 1068, the NLS2 J1203+1624, and another 291 SDSS type-2 candidate in Zakamska et al. (2003) are also included in Figure 3\footnote{Note that Mrk 1126 is absent in the color-color diagram because it is not covered in the SDSS database. And NGC 1068 is so bright and extended, the PSF magnetude is adopted in the $g$ band rather than the petrosian magnitude.}. With the largest $g-W_4$ and $g-J$ colors, Mrk 1239 is clearly the ``reddest'' object among the prototype NLS1s in Osterbrock \& Pogge (1985).

As a type-1 AGN, the IR flux in Mrk 1239 is relatively too strong as compared with the UV-optical radiation.
The integrated AGN luminosity in the rest-frame IR is $L_{[1-20]\micron} = 3.6\times 10^{44} \rm erg\ s^{-1}$, nearly 10 times the AGN UV-optical luminosity\footnote{we extrapolate the $GALEX$ fluxes into the EUV.} of $L_{[0.01-1]\micron} = 3.8 \times 10^{43} \rm erg\ s^{-1}$. Usually referred to as reprocessed emission of dusty torus, the IR radiation from the quasar should be no stronger than the incident UV-optical radiation from the accretion disk. The only explanation is that Mrk 1239 is heavily obscured, i.e. the actual UV-optical radiation that illuminates the torus is much stronger than we have observed.
This result is similar to the polarimetric observations that Mrk 1239 is very different from typical type-1s. On the other hand, similar to J1203+1624 and NGC 1068, the $g-W_4$=12.35 color of Mrk 1239 is significantly larger than most of the NLS1s in Zhou et al. (2006). And large $g-W_4$ values are commonly seen among type-2s as shown in Figure 3(b). Such a large optical-MIR color in Mrk 1239 and type-2s suggest that the optical flux in these targets are significantly lower than in normal type-1s. This is usually caused by dust reddening, which tend to extinct radiation more strongly at shorter wavelengths. The high degree of polarization and very red optical-IR colors in Mrk 1239 are clearly contradicting its original identification as a type-1 AGN, making it an extremely interesting object in the framework of the AGN unifcation. On the other hand, the $g-J$ color of Mrk 1239 is also significantly larger than normal type-1s and even typical type-2s. All these properties are discussed in detail in \S7.

\begin{figure}
\includegraphics[width=0.49\textwidth]{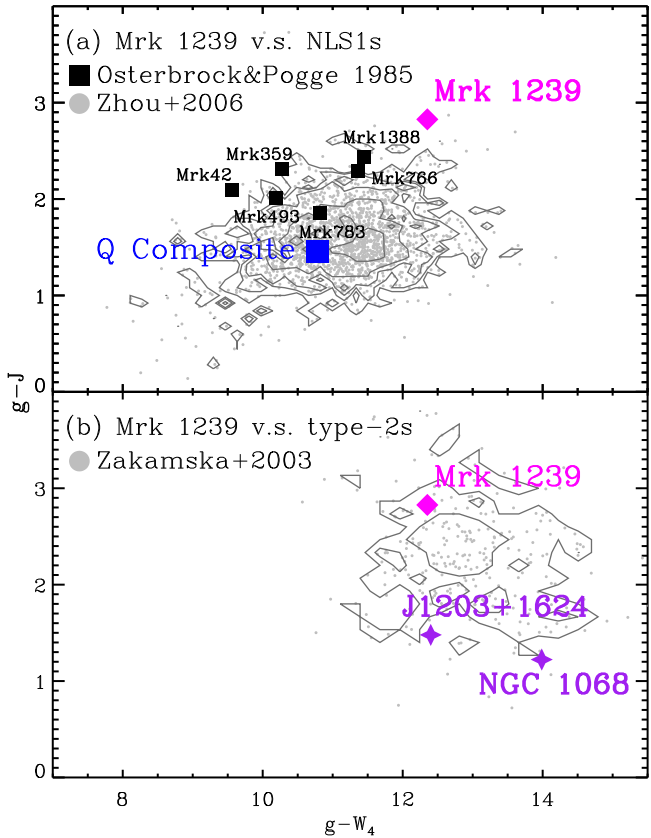}
\caption{
$g-J$ vs. $g-W_4$ color of Mrk 1239 is compared with typical NLS1s in panel (a), and with typical type-2s in panel (b). Even though it was originally identified as a NLS1, Mrk 1239 deviates from typical NLS1s, and on the contrary resembles typical type-2s.}
\end{figure}

\section{SED Decomposition}
The optical-IR images of Mrk 1239 clearly show an extended structure, indicating the presence of prominent host starlight. The high degree of polarization of Mrk 1239 indicates that there is a significant fraction of scattered light (\S3). And the red optical-infared color suggests that a reddened component dominates in the IR (\S4).  To quantify the proportion of these three contributions, the following decomposition process is introduced.

\begin{figure}
\includegraphics[width=0.49\textwidth]{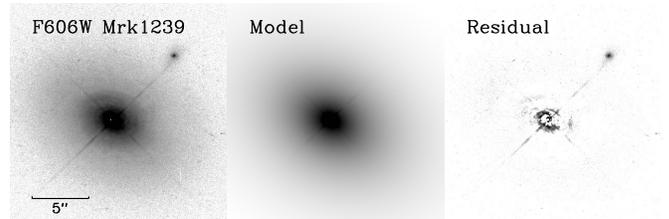}
\caption{Decomposition of $HST$/WFC3-pc1 image of Mrk1239 (left panel) into a simple PSF+S\'ersic model (middle). The residual image is in the right panel.}
\end{figure}

The high-resolution $HST$/F606W image of Mrk 1239 enables a clear separation of the point-like nucleus from the extended host galaxy. A simple PSF (for the nucleus) + S\'ersic (1968) (for host galaxy) model can fit the data well with GALFIT (Peng et al. 2002). 
A comparison between the data and best-fit results is shown in Figure 4 (a detailed description can be found in \S A1). The resultant profile parameters of $HST$/F606W image are then applied to fit the SDSS and UKIDSS images of Mrk 1239 and the fluxes of both the PSF component and the S\'ersic component of SDSS and UKIDSS bands are obtained.

The SED of the S\'ersic component (cyan squares in Figure 5) can be approximated with a simple stellar population (SSP) template (Bruzual \& Charlot 2003) at an age of $\sim$ 2.5 Gyr (cyan line in Figure 5). 

The PSF flux decreases dramatically from the mid infrared down to the optical band. And the flux-decreasing then stops at SDSS-$i$ band (around 0.8 $\rm \mu m$ in the rest-frame) and flattens down to the UV. Such a flattening in the optical-UV is unlikely caused by reddening because extinction curves with a turning point at around 0.8 $\rm \mu m$ is hardly seen in the literature. Such behavior is similar to the three reddened NLS1s reported by Zhang et al. (2017). Following similar treatment as  Zhang et al. (2017), the quasar composite ($Q_\lambda$) is assumed as the absorption-free template. The template is then reddened with the extinction curve of Small Magellanic Cloud (SMC; $A_\lambda^{\rm SMC}(E_{B-V})$; Prevot et al. 1984) in the form $a\ 10^{{\rm -}0.4 A_\lambda^{\rm SMC}(E_{B-V})}$ $Q_\lambda$ to fit the PSF flux of Mrk 1239, where the two parameters are the scaling factor $a$ and the color excess \ebv.
When doing the fitting, only the IR photometries ($\lambda > 1$\micron) which are dominated by transmitted radiation, are taken into account. The color excess is found to be \ebv=1.6$\pm$0.1.

By applying the best-fit color excess of \ebv=1.6$\pm$0.1, the observed PSF IR fluxes can be well described by the model (pink line in Figure 5). At such a high extinction, transmitted AGN radiation is predicted to decrease dramatically at shorter wavelengths and goes gradually negligible in the optical and UV.

As shown in Figure 5, after removing the unpolarized starlight, the polarized flux is found to be $\sim$1/7 of the PSF component, indicating a high intrinsic $P$ value of $>$10\%. As discussed in detail by Antonucci (1984), such a high $P$ in the optical band is most likely originated from electron scattering processes.
It is also found that the PSF components’ $g - r$ color (purple squares in Figure 5) is nearly the same as the polarized flux (blue dots in Figure 5) of Mrk 1239, suggesting that the intrinsic $P$ value does not change dramatically with wavelength, consistent with typical electron scattering.
A similar broad Balmer decrement for both the polarized(\halpha/\hbeta$\approx$5.3) and the total (\halpha/\hbeta$\approx$4.8) flux spectra is also found, consistent with the scenario. Considering the average \halpha/\hbeta ratio of 3.06 for blue AGNs (Dong et al. 2008), the Balmer decrement of $\sim5$ for the scattered AGN radiation corresponds to an extinction of $\sim$0.5 if SMC extinction curve is applied. The result is similar to the extinction of \ebv$\approx$0.54 estimated by Rodr{\'{\i}}guez-Ardila \& Mazzalay(2006) based on comparing continuum of Mrk 1239 with Akn 564 as an extinction-free template.

As a brief summary, three components are found in the UV-optical-IR SED of Mrk 1239: the host galaxy starlight with an SSP age of $\sim$ 2.5 Gyr, a scattered AGN radiation with $P>10$\% in the optical band, and a reddened AGN radiation with an \ebv of 1.6. Based on the extinction-free model (grey line in Figure 5.) for the transmitted AGN radiation, the luminosity at rest frame 5100\AA is estimated to be 2.5$\times 10^{44}\rm erg\ s^{-1} cm^{-2} \AA^{-1}$. A bolometric luminosity of $L_{\rm bol} = 2.5 \pm 0.5 \times10^{45} \rm erg\ s^{-1}$ is then obtained assuming its scaling relation with $L_{5100}$ (Richards et al. 2006).
If the extinction-free SED of Mrk 1239 resembles the broadband AGN template of Mathews \& Ferland 1987, an incident rate of the ionizing photon is then $Q_{\rm H} = 2.6\pm0.5 \times 10^{56}\rm\ s^{-1}$. If an efficiency factor of 0.1 is assumed, the mass accretion rate is then $\dot{M}({\rm accretion}) = 4.5\pm0.9 \times10^{-3} M_\odot/\rm yr^{-1}$.
By measuring the NIR BELs, the width of the transmitted hydrogen BELs is $FWHM_{\rm IR~BEL} = 1090 \pm 80$ \kms (\S A2). The mass of the supermassive black hole (SMBH) is then $M_{\bullet} = 9.4 \pm 0.6 \times 10^6 M_\odot$, if its correlation with the BEL width and the luminosity is assumed (Greene \& Ho 2005). Based on the estimated bolometric luminosity, the Eddington ratio is then 2.2$\pm0.6$. In very luminous AGNs ($L_{\rm bol} > 10^{45} \rm erg\ s^{-1}$) such as Mrk 1239, the efficiency factor can be significantly larger than 0.1 as suggested by the detection of strong jet power in them (e.g. B{\"a}r et al.2019). Thus the Eddington ratio of Mrk 1239 could be a bit lower than the estimated value of around 2.2, but is still very likely close to 1. Finally, based on modeling of the S\'ersic fluxes, the luminosity of starlight is $L_* \approx 4.5\times10^{43}\rm erg\ s^{-1}$, corresponding to a stellar mass of $M_* \approx 1.5\times10^9 M_\odot$.

\begin{figure*}
\includegraphics[width=0.99\textwidth]{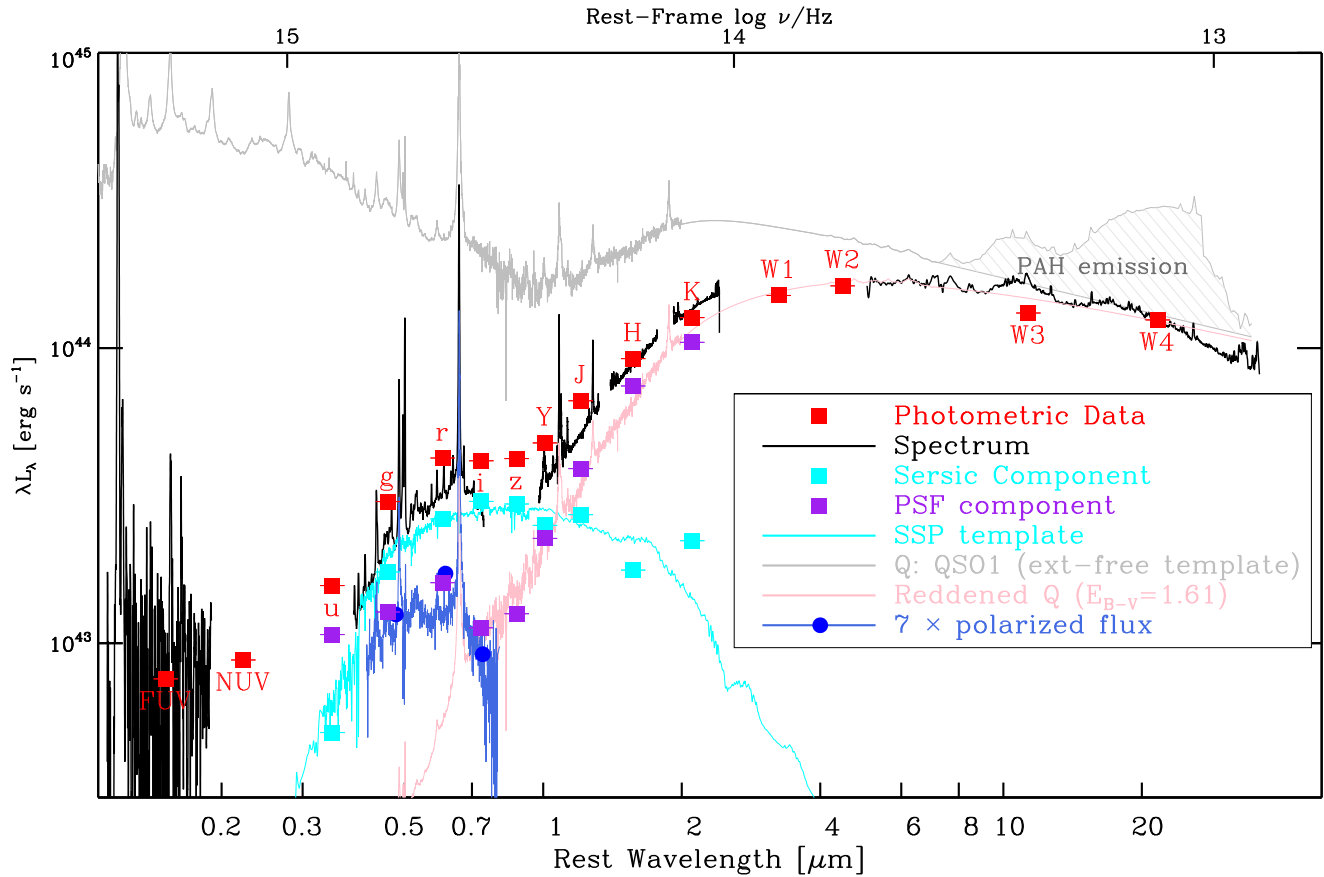}
\caption{
SED decomposition of Mrk 1239 is shown. The black line indicates the observed spectra. From left to right, they are $IUE$ FUV spectrum, 6dF optical spectrum, P200/TripleSpec NIR spectrum, and $Spitzer$/IRS spectrum in the MIR. Red squares denote flux of photometric data. SED of the QSO composite is displayed in gray as a comparison.
SDSS and UKIDSS photometric data are decomposed into S\'ersic (cyan) and PSF (purple) components.
The S\'ersic (cyan) component is fitted with a SSP template(sky blue line). In the optical band, the PSF component is found to be similar to polarized flux (blue lines and dots), in both their $g-i$ color, and the broad Balmer decrement. The PSF component in the IR ($\lambda > 1\micron$) is best-fitted with the quasar template, reddened at an \ebv of 1.6 (pink line).}
\end{figure*}

\section{Emission and Absorption Line Spectrum}

\begin{figure}
\includegraphics[width=0.49\textwidth]{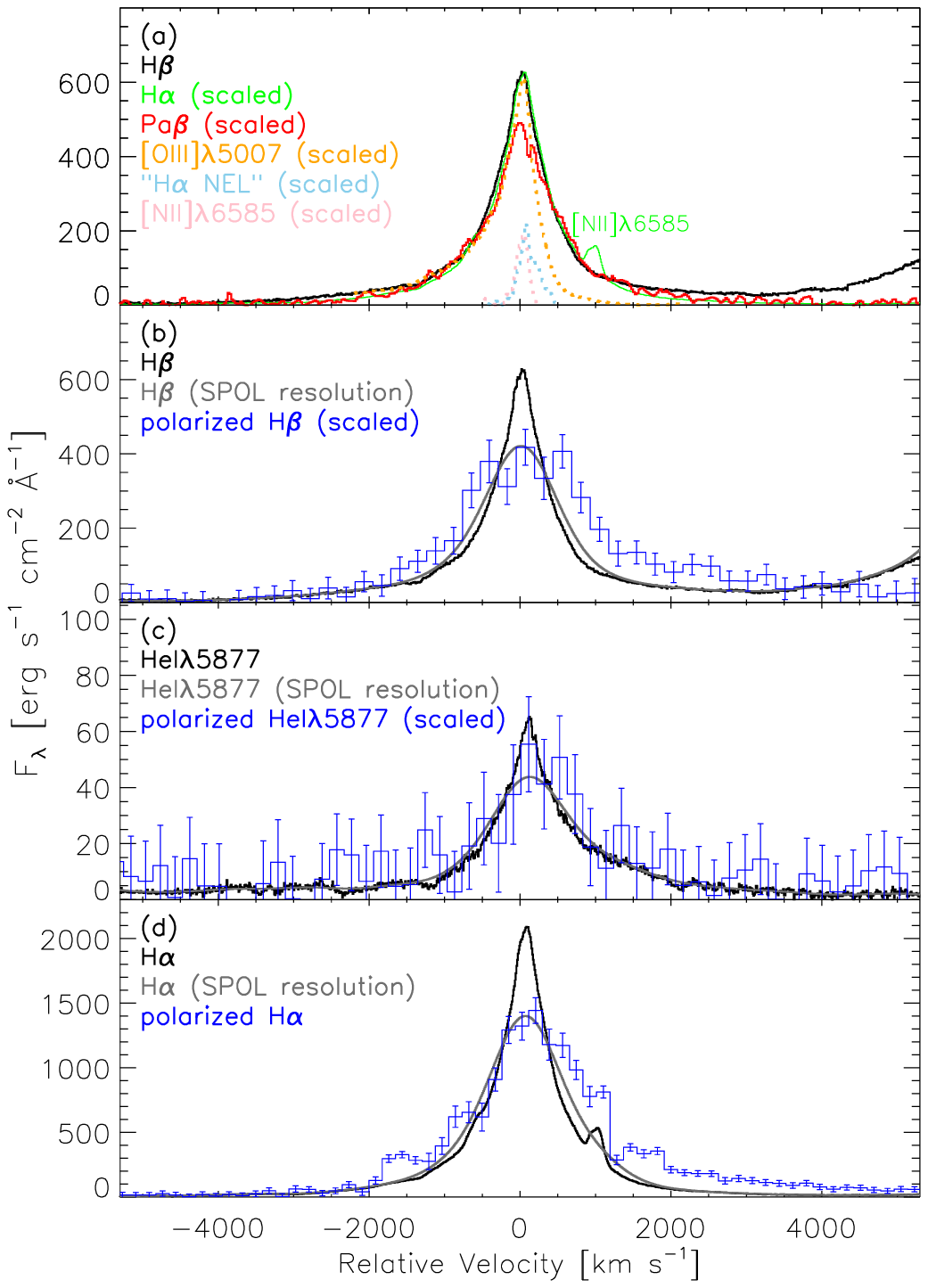}
\caption{(a) Comparison of profiles of different emission lines in their common velocity space. After subtracting the scaled \pabeta profile, the ``\halpha NEL'' component (blue dotted line) and \nii$\lambda$ 6585 (pink dotted line) are obtained. Line profiles of the total flux (from ESI data, black lines), total flux at SPOL resolution (grey lines) and polarized fluxes (blue lines) of \hbeta, \hei$\lambda$5877, and \halpha are compared in Panel (b), (c), and (d) respectively.}
\end{figure}

At an estimated Eddington ratio of around 2.2, the nucleus of Mrk 1239 is highly active and luminous. Similar to luminous quasars like 3C 273, the broad emission lines should outshine the narrow emission line components if Mrk 1239 is free from dust obscuration. We thus expect that at longer wavelengths where extinction is relatively weak, the Paschen emission line should be dominated by the transmitted broad emission line component. This is indeed found by comparing \pabeta with Balmer lines in Figure 6(a). \halpha and \hbeta are found to have the same emission line profiles, indicating that bulk fluxes of Balmer emission lines have similar physical origins, i.e. coming from scattering as predicted in the SED analysis. On the other hand, contribution from the broad emission line component in \pabeta is found to be relatively stronger, since its narrow peak at the line core appears to be missing as compared with Balmer emission lines. After subtracting the scaled \pabeta emission line from \halpha, the profile of the ``\halpha NEL'' component and \nii$\lambda$ 6585 are found to be similar in the core region. Comparison between the profile of Balmer and Paschen emission lines yields that the obscuration in broad emission lines is significantly stronger than in narrow emission lines. The physical scale of the obscuring clouds is then significantly larger than the broad emission line region, but comparable or smaller than that of the narrow emission line region (NELR $\leqslant$100 pc to the SMBH, Peterson 1997). In addition, SED modelling suggests that the continuum in the NIR, which is usually regarded as emitted by hot dust grains in the inner surface of the torus (Barvainis 1987), endures prominent extinction as compared with the type-1 quasar template. Then the distance and scale of the obscurer should be larger than the sublimation radius of $R_{\rm sub} = 0.6$pc of Mrk 1239 according to Barvainis (1987).

\begin{figure}
\includegraphics[width=0.49\textwidth]{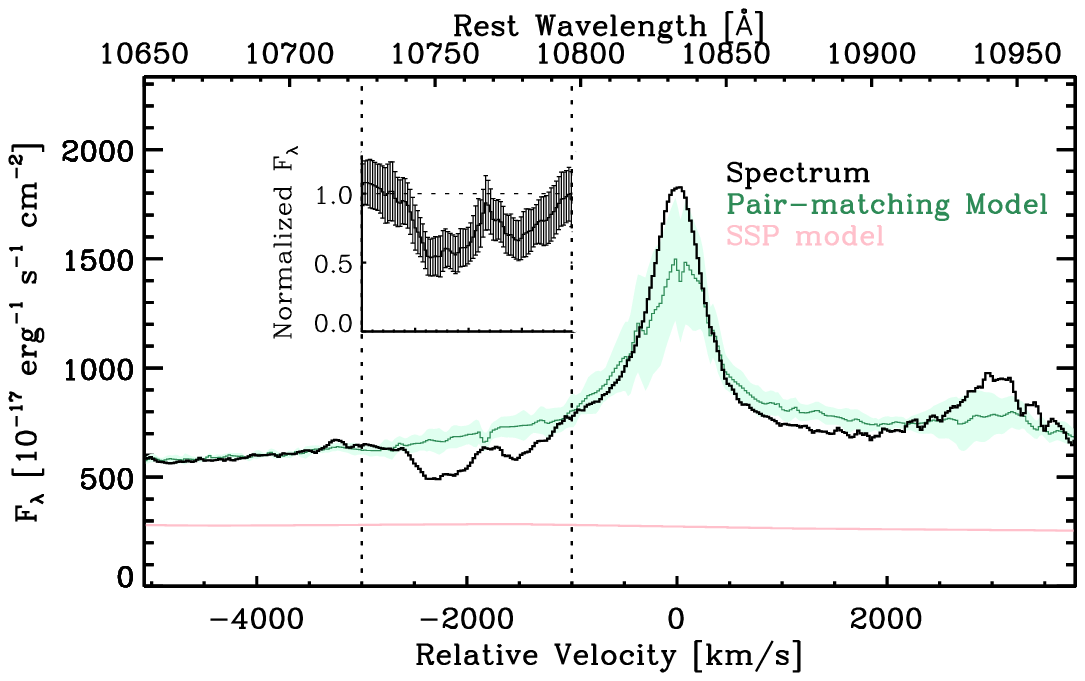}
\caption{The spectrum at around \hei*$\lambda$10830 is shown. The pair-matching absorption-free model is in the green line, and the SSP model derived in SED modeling is displayed with the pink line. The normalised flux density in the velocity range of [-3000,-1000]\kms is shown in the insert panel.}
\end{figure}

The extinction (\ebv=1.6$\pm$0.1, \S4) of the transmitted AGN radiation corresponds to a hydrogen column density of log $N_{H}/\rm cm^{-2}$ = $22.0\sim22.9$  if dust-to-gas ratios of the metal-poor SMC (Prevot et al. 1984) and the metal-rich Milky Way (Bohlin et al. 1978) are considered. This value is lower than the column density $log N_{\rm H}/\rm cm^{-2} = 23.5_{-0.3}^{+0.2}$ of the X-ray absorber in Mrk 1239 (Grupe et al. 2004), but the difference of the inferred column densities is not significantly large if the uncertainties are considered. 
Prominent absorption lines should be found for absorbers at such a high column density. We do find significant \hei*$\lambda$10830 BAL in the infrared, where the flux is dominated by transmitted radiation from the AGN. Because the emission of \hei*$\lambda$10830+\pagamma is complex, we apply the pair-matching method in obtaining the absorption-free model in this spectral region. Ninety-Two infrared quasar spectra from infrared quasar spectral atlases (Glikman et al. 2006, Riffel et al. 2006; Landt et al. 2008) are applied as templates. Even though the absorption-free model fails to reproduce the emission line peaks of Mrk 1239, the spectral region around the \hei*$\lambda$10830 BAL trough ($\sim$[-3000,-1000]\kms) is approximated reasonably well. Because there is no evidence that the host galaxy is obscured, we assume that the starlight is not covered by the absorber, and remove the flux of the SSP model (pink solid line in Figure 7) before calculating the normalised flux (inner panel of Figure 7). The covering faction is probably large, because the SED of the PSF component in the NIR appears to be dominated by reddened AGN radiation as shown in Figure 5.
A lower limit of the column density is found to be log $N_{\rm HeI (2^3S)}/\rm cm^{-2}$ = 13.5 $\pm$ 0.2, by integrating the apparent optical depth (AOD, Savage \& Sembach 1991) of the \hei BAL. In the mean time, no apparent absorption is found in the speed range of [-3000,-1000]\kms around \halpha despite the fact that a part of the AGN flux around \halpha comes from transmitted radiation. Based on the spectra around \halpha in this velocity range, an upper limit of column density for the neutral hydrogen at the excited state $n$=2, log $N_{\rm HI n=2}/\rm cm^{-2} <$ 13.1, is obtained. Given that the \hei absorption line is blueshifted at $\sim 2000$\kms relative to the systemic redshift, the absorber is very likely originated from the AGN outflows. Based on the column density measurements, we run CLOUDY (c17.02; Ferland 1998) simulations to investigate the photoionization of the obscuring outflow gas (\S A3). The results indicate that the distance of the dusty outflow to the central black hole is in the range 0.01--1 kpc, consistent with the qualitative estimates inferred from the obscuration of each spectral components.

The detection of prominent scattering indicates that large amounts of scattering clouds posit around the nucleus of Mrk 1239. The high value of $P$ indicates that the scatterer covers a significant fraction of the AGN radiation. Similar to Goodrich et al. (1989), we find that $P$ decreases as wavelength increases in Mrk 1239, indicating that the scatterer can be dusty because the scattering cross section of dust grains generally decreases as wavelength increases. A dusty scatterer is also favoured because the scattered hydrogen Balmer emission lines are at a similar width to the directly observed (transmitted) Hydrogen Paschen emission lines (Figure 6). Otherwise, scattering by electron will tend to broaden the line width (the broadened width is 1300 \kms at a typical temperature of $T_e = 10,000$K, Goodrich et al. 1989), and scattered Hydrogen Balmer emission lines will be significantly broader than the observed width of just $FWHM_{\rm opt\ BEL} =  830\pm 10$ \kms. Because the polarization of NELs is different from the continuum and BELs, the scatterer of the continuum and BELs should be at relatively small scales. Otherwise, a distant scatterer at scales larger than the narrow emission line region will polarize the NELs in ways similar to the continuum and the BELs. Similar to the previous spectropolarimetries of Mrk 1239 (Goodrich et al. 1989), we also find red wings in the polarized broad emission lines like \hbeta, \hei$\lambda$ 5877, and \halpha (Figure 6(b)-(d)), indicating that the scattering clouds are also flowing outwards from the AGN. As a brief summary, we find that the scatterer is dusty and outflowing at a distance $\leqslant 100$ pc, all of these properties resemble the dusty absorber been discussed. The presence of both the obscurer and the scatterer indicates the existence of large amounts of outflowing clouds at close distances to the central SMBH. On the other hand, a blueshifted wing centered at $\sim$ 1000\kms is also found in the asymmetric \oiii$\lambda$5007 emission line (Figure 6(a)), confirming the existence of outflows in Mrk 1239.
If the obscurer and the scatterer are correlated, some of their properties, i.e. shifting speeds ($v \sim$ -2000 \kms), column densities (10$^{22\sim23.5} \rm cm^{-2}$), distances ($R\in[ 0.01,0.1]$kpc), and global covering fraction ($\Omega\sim 10\%$), can also be similar. Following Borguet et al. (2012), the upper limits of mass-flow rate ($\dot{M}({\rm outflow})$) and kinetic luminosity ($\dot{E}_k$) are then,
\[\dot{M}({\rm outflow}) \leqslant 0.31 M_{\odot} \rm yr^{-1},\]
\[\dot{E}_k({\rm outflow}) \leqslant 10^{41.6} {\rm erg\ s^{-1}} = 10^{-3.8} L_{\rm bol}\].
The obscurer and the scatterer thus indicate that there can be powerful outflows at close distances in Mrk 1239.

\section{Mrk 1239: a Type-2 Counterpart of NLS1?}

The first reports of optical polarization in Seyfert nuclei can be dated back to the 1960s (e.g. Visvanathan \& Oke 1968; Walker 1968), especially in NGC 1068. Angel et al. (1976) then conclude that the significant optical polarization ($P \sim$ 5\%) in NGC 1068 is originated from scattering of nuclear light by dust grains. After subtracting the unpolarized starlight from the host galaxy, a very high degree of polarization of 16\% in NGC 1068 is revealed for both the continuum and broad emission lines perpendicular to the nuclear symmetry axis determined by the radio morphology. Similar to NGC 1068, high degree of polarization at perpendicular to the radio structure is found to be universal in Seyfert-2s (Antonucci 1983). 
A schematic cartoon of typical type-2s like NGC 1068 is displayed in Figure 8. In this cartoon, although the nucleus (accretion disk and BELR) is obscured (in the UV-optical band) and reddened  (in the IR) by clouds in the toroidal structure at the equatorial plane, clouds in the polar region can scatter a significant portion of light into our sight. The transmitted AGN in the IR then leads to a red optical-IR color. And the scattered component in type-2s usually shows significant polar scattering, which can be revealed with polarimetric observations (Antonucci 1983).
The first NLS2 reported of having broad Balmer emission lines with $FWHM<2000$\kms, i.e. J1203+1624, also shows a high degree of polarization of ($P \sim$ 7.3\%). After the removal of starlight, the intrinsic polarization of the nucleus is around 13.6\%, similar to NGC 1068. Spectropolarimetry is undoubtedly a powerful probe to hidden nucleus of type-2s. As introduced in \S3, the directly observed $P$ of Mrk 1239 is at a similar level to typical type-2s. And the intrinsic $P$ value of Mrk 1239 is also at a similar level as compared with NGC 1068 and J1203+1624. Modeling of line spectra of Mrk 1239 shows that the transmitted broad emission lines in the IR ($FWHM_{\rm IR~BEL} = 1090 \pm 80$ \kms) are considerably broader than the scattered BELs in the optical band ($FWHM_{\rm opt\ BEL} =  830\pm 10$ \kms, see \S A2 for a detailed description). This result seems controversial because the scattering processes tend to broaden emission lines by the thermal motion of the scattering particles. However, as shown in the cartoon of Figure 8, the transmitted AGN component (magenta line) is likely viewed at large inclinations from face-on as compared with the scattered AGN radiation. At large inclinations, the projected velocity of the Keplarian rotation of the disk-like BELR will also be stronger. Thus the transmitted emission lines can be broader than the scattered emission lines from smaller inclinations. However, NIR spectroscopy with higher quality for Mrk 1239 and its analogs is needed for a more detailed comparison study of emission line profiles in type-2 AGNs. In the radio band on the other hand, the high-resolution Very Long Baseline Array (VLBA) image of Mrk 1239 yields a two-sided jet-like structure within 40 pc, at a position angle of 47$^\circ$ (Doi et al. 2013), nearly perpendicular to the scattering position angle of 131$^\circ$. Such a high degree of polarization and polar scattering is strong evidence that Mrk 1239 could be a type-2 AGN.

\begin{figure*}
\includegraphics[width=0.99\textwidth, trim=0 0 430 0,clip]{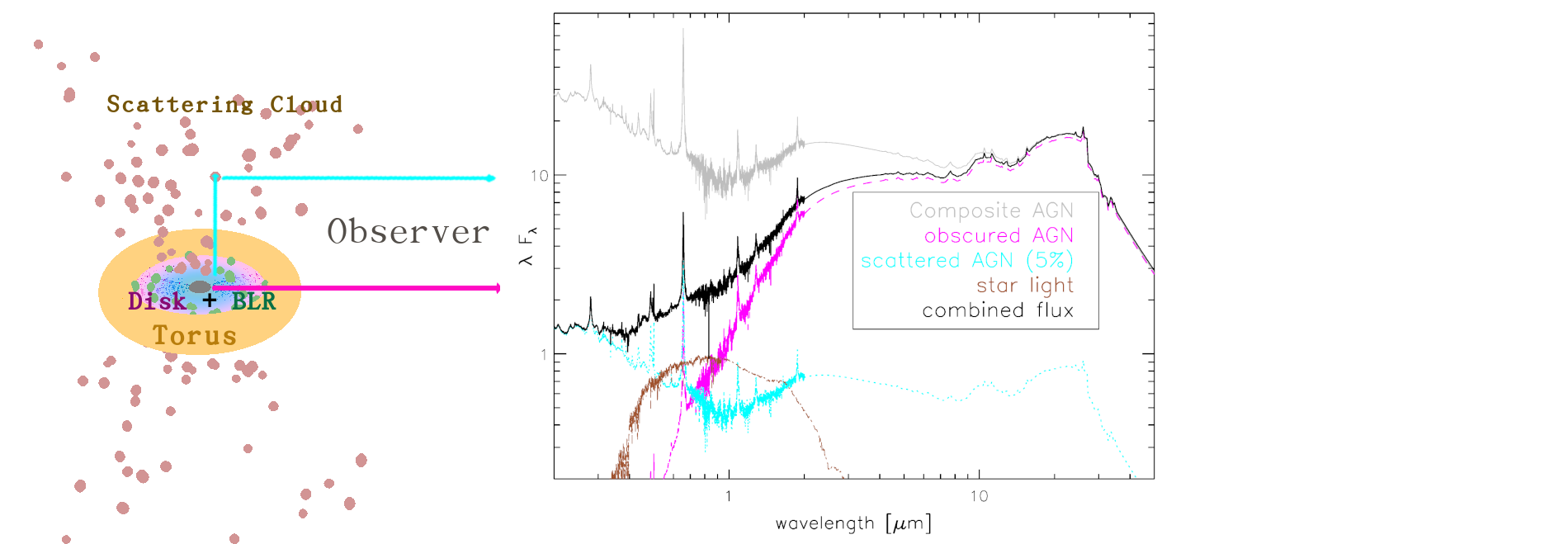}
\caption{A schematic cartoon of typical type-2s such as NGC 1068 is shown at left, with SED models for different components displayed in the right panel. Besides an obscured and reddened light path (magenta line), a significant portion of the light from the nucleus (disk + BELR) can be scattered by clouds in the polar region into our sight line (cyan line).}
\end{figure*}

If Mrk 1239 is a type-2 counterpart of NLS1, prominent obscuration to the nucleus is expected. By analysing the UV-optical-IR SED, a heavily reddened (\ebv$\approx$1.6) and transmitted AGN component is revealed that dominates in the IR. And the optical fluxes of Mrk 1239 can be scattering dominated as it is similar to the polarized flux spectrum in both the continuum slope (a similar $g-i$ color) and  Balmer decrement. The scattered AGN radiation is also mildly obscured at \ebv$\sim 0.5$. The analysis of X-ray data of Mrk 1239 also reveals two components in Mrk 1239, a deeply obscured one (log $N_{\rm H}/\rm cm^{-2} = 23.5$) that dominates in the hard X-ray ($E > 3$eV), and a scattered component with mild absorption (log $N_{\rm H}/\rm cm^{-2} = 20.8$) in the soft X-ray (Grupe et al. 2004). As for the foreground absorption for the scattered component in X-ray, if typical dust-to-gas ratios of the Milky Way and SMC are assumed, the corresponding extinction is at the level of \ebv=0.01--0.1, at similar levels of the \ebv$\sim$0.5 found for the scattered light in the optical band.

Under the framework of the unified model of AGNs, type-2 AGNs are viewed at large inclinations from face-on and obscured by dusty clouds at the equatorial plane. The steep spectrum and jet-like structure in the radio band suggest that the nucleus of Mrk 1239 is very likely viewed at large inclinations (Ulvestad et al. 1995; Condon et al. 1998; Doi et al. 2013). As discussed in the previous sections, the nucleus of Mrk 1239 appears to be obscured by dusty clouds at distances in between the sublimation radius and the NELR, which is similar to the physical scale of the dusty torus. Elitzur \& Shlosman 2006 propose that the dusty torus can be clumpy outflowing clouds driven from the disk. Given that the obscuring clouds are outflowing at $\sim2000$\kms, the ``dusty torus'' of Mrk 1239 might also be entangled with the global outflowing clouds.

\section{Implications and Future Works}
After a detailed comparison with typical NLS1s, one of the 8 prototypes, Mrk 1239 is found to be a peculiar outlier, having much more in common with type-2 AGNs than with NLS1s. The broad band data, from X-ray, through UV-optical-IR, to the radio, can all be consistently explained if Mrk 1239 is a type-2 counterpart of NLS1s. The process of reviewing the classification of Mrk 1239 also leads to intriguing facts about this amazing target. After decomposing the SED into starlight, scattered AGN radiation, and transmitted AGN radiation, properties of both the nucleus ($L_{\rm bol} = 2.5 \pm 0.5 \times10^{45} \rm erg\ s^{-1}$, $M_{\bullet} = 9.4 \pm 0.6 \times 10^6 M_\odot$, $L_{\rm bol}/L_{\rm edd} = 2.2\pm0.6$) and the host galaxy ($L_* \approx 4.5\times10^{43} \rm erg\ s^{-1}$) are explored, which is usually hardly achieved for AGNs at super-Eddington accretion rates. Based on analysing polarized/total flux of continuum and emission/absorption lines, Significant amounts of outflowing and scattering clouds are revealed at physical scales between the BELR and the NELR, similar to the obscuring clouds. The high-angular resolution spectroscopy and photometry in the future will help to confirm the classification of Mrk 1239, and explore the abundant circum-nucleus contents in it.

The study of J1203+1624 in our previous work indicates the existence of type-2 conterparts of NLS1s, i.e. NLS2s. 
As one of the eight prototype NLS1s, if Mrk 1239 is intrinsically a type-2 as suggested by all the evidences across the wavelengths, a substantial amount of previously identified NLS1s can be obscured. At a glance of the NLS1 samples, 79 out of 1946 NLS1s in Zhou et al. (2006) are found to have optical-MIR colors larger than Mrk 1239 $g-W_4 > 12.35$. Obscuration and scattering can widely exist in these sample targets. Follow-up observations and analysis in our future works will reveal the hidden nuclei, circum-nuclei gas contents, and host galaxies of these obscured NLS1s, and their implications to the AGN unification and evolution will be explored.

\acknowledgments
We appreciate the detailed and constructive comments from an anonymous reviewer, which improve greatly the quality of this paper in data analysis, discussion, material organizing, and English writing.
This work is supported by National Natural Science Foundation of China (NSFC-11903029). X. S is supported by Shanghai Natural Science Foundation (20ZR1463400). P. J is supported by National Natural Science Foundation of China (NSFC-11973037).
H. L, X. P, and L. S are supported by Natural Science Foundation of Anhui (1808085MA24). W. L acknowledges support from the Natural Science Foundation of China grant (NSFC-11703079) and the "Light of West China" Program of Chinese Academy of Sciences (CAS).  We acknowledge the use of the Multiple Mirror Telescope at Fred Lawrence Whipple Observatory
and the Hale 200-inch Telescope at Palomar Observatory through
the Telescope Access Program (TAP), as well as the archival data from the $GALEX$, SDSS, 2MASS, UKIDSS, $WISE$ Surveys and Spitzer space telescope.
TAP is funded by the Strategic Priority Research Program. The Emergence of Cosmological Structures
(XDB09000000), National Astronomical Observatories, Chinese Academy of Sciences, and the Special Fund
for Astronomy from the Ministry of Finance. Observations obtained with the Hale Telescope
 were obtained as part of an agreement between the National Astronomical Observatories, Chinese
Academy of Sciences, and the California Institute of Technology. The WFPC2 image (observed by NASA/ESA $Hubble Space Telescope$) and FUV spectra (observed by $IUE$ satellite) were obtained from the data archive at the Space Telescope Science Institute. STScI is operated by the Association of Universities for Research in Astronomy, Inc. under NASA contract NAS 5-26555. This research has also made use of the Keck Observatory Archive (KOA), which is operated by the W. M. Keck Observatory and the NASA Exoplanet Science Institute (NExScI), under contract with the NASA.

\appendix

\section{A1. Image Decomposition}

We start the decomposition of Mrk 1239's images into AGN and host galaxy components with the $HST$/F606W image. By convention, it is assumed that the AGN is in a PSF profile in the image, and the host galaxy follows S\'ersic distribution ($r^{1/n}$). The synthetic PSF model generated using the TinyTim software (Krist 1995) is adopted, which is found to be reasonable approximations by both simulations (Kim et al. 2008) and empirical works (Jiang et al. 2013). It is found that a simple GALFIT (Peng et al. 2002) modeling with the PSF+S\'ersic model can approximate the image reasonably well (Figure 4). Best-fit results (mid-panel of Figure-1) suggests a S\'ersic index of $n=4.03\pm$0.02, an effective radius $R_e = $3\arcsec.72$\pm$0\arcsec.02 , an axis ratio $b/a = 0.76\pm0.01$, and a position angel of $PA$=57\degree.7$\pm$0\degree.1 for the host galaxy. 

At a lower resolving power, the decomposition of SDSS and UKIDSS images (listed in Figure 9) is relatively more difficult, especially at the blue (SDSS $u$) and the red (UKIDSS $J$, $H$, $K$) end, where the images of Mrk 1239 appear point-like, and the extended host galaxy becomes weak and hard to be recognized or modeled. We thus fix the parameters of the S\'ersic component based on the $HST$/F606W image in the fitting of SDSS and UKIDSS images. In all the SDSS and UKIDSS images of Mrk 1239, the images of 3 reasonably bright nearby stars are stacked to obtain the PSF model. In the  UKIDSS $H$ $K$ bands, saturated pixels at the center of Mrk 1239 are masked. The fluxes of both the S\'ersic and PSF components in each SDSS and UKIDSS bands are then derived (plotted in Figure 5), and typical uncertainties of the fluxes are at the level of $\sim~4\%$.

\begin{figure*}
\includegraphics[width=0.99\textwidth]{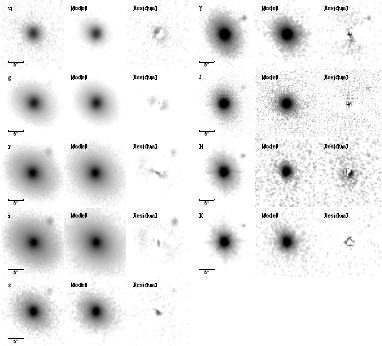}
\caption{The observed images, best-fit GALFIT models, and residuals in the SDSS (left three columns) and UKIDSS (right three columns) bands are displayed.}
\end{figure*}

\section{A2. Emission Line Modeling}
Following Zhou et al (2006), we carry out an iterative
approach to  fit both the continuum and emission lines of the 
6dF optical spectrum of Mrk 1239. 
Here is a brief summary. Step 1, the  optical continuum  is  fitted  in
continuum windows with a combination of host starlight and 
featureless continuum from the nucleus. The simple stellar population (SSP) models in Bruzual \&
Charlot (2003) are  assumed as host starlight, reddened with a
Milky Way (MW) extinction law (Fitzpatrick \& Massa 2007). The
nuclear emission includes a power-law (PL) continuum and a  two-component
analytic \feii emission model (Dong et al. 2005) which incorporates the
VJV04 \feii template (V{\'e}ron-Cetty et al. 2004). During the continuum fitting, 
spectral regions around strong stellar absorption features (grey regions in 
the lower panel of Figure 10) are weighted higher than the rest of the data points so as to ensure a reasonable fit to of the SSP component.
Step 2, to obtain the observed line spectrum, the fitted continuum model 
in the previous step is subtracted. A high-order polynomial is then applied to subtract the residual
fluxes after continuum subtraction, so as to minimize the contamination of continuum modeling to the fit of the line spectrum.
Besides the prominent NELs, a broad base is clearly seen at \halpha, 
indicating the presence of a (low-ionization) BEL component. A test fitting of the \oiii+\hbeta
complex shows that forbidden lines can be fitted with at least two Gaussians, a narrow Gaussian for the regular narrow emission line (NEL) component, and a relatively broad Gaussian for the broad base. In the mean time, a Lorentzian profile can approximate the Balmer broad emission lines (BELs) better than a Gaussian profile.
Thus, each of the identified forbidden or permitted lines in the line spectrum is assigned a 2-Gaussians or Lorentzian profile, with their line centroids and widths bound in the velocity space. Then the line spectrum is modeled. Step 3, repeat the fitting of continuum and emission lines until model parameters converge.
The converged continuum and emission line models are displayed in Figure 10.
The systematic redshift of Mrk 1239 is then determined by the SSP component, which is $z_{\rm SSP} = 0.01980\pm0.00003$. The centroid of both the  component ($FWHM_{\rm opt~BEL} =  830\pm 10$ \kms) and the narrow NEL Gaussian ($FWHM_{\rm NEL} = 490 \pm 10$) is close to the systematic redshift, with shifting speeds of $-10 \pm 1$ \kms and $-12 \pm 1$ \kms respectively. In the mean time, the relatively broad base ($FWHM_{\rm NEL~Base}  = 1330 \pm 10$\kms) is discovered in high-ionization forbidden lines like \oiii double, \fev $\lambda$ 4072, \fevii $\lambda\lambda$ 5722, 6088, and \fex $\lambda$ 6376. The centroid of the NEL broad base is blueshifted at a speed of $420\pm10$\kms, indicating the presence of emission line outflows in Mrk 1239. 

\begin{figure*}
\includegraphics[width=0.99\textwidth]{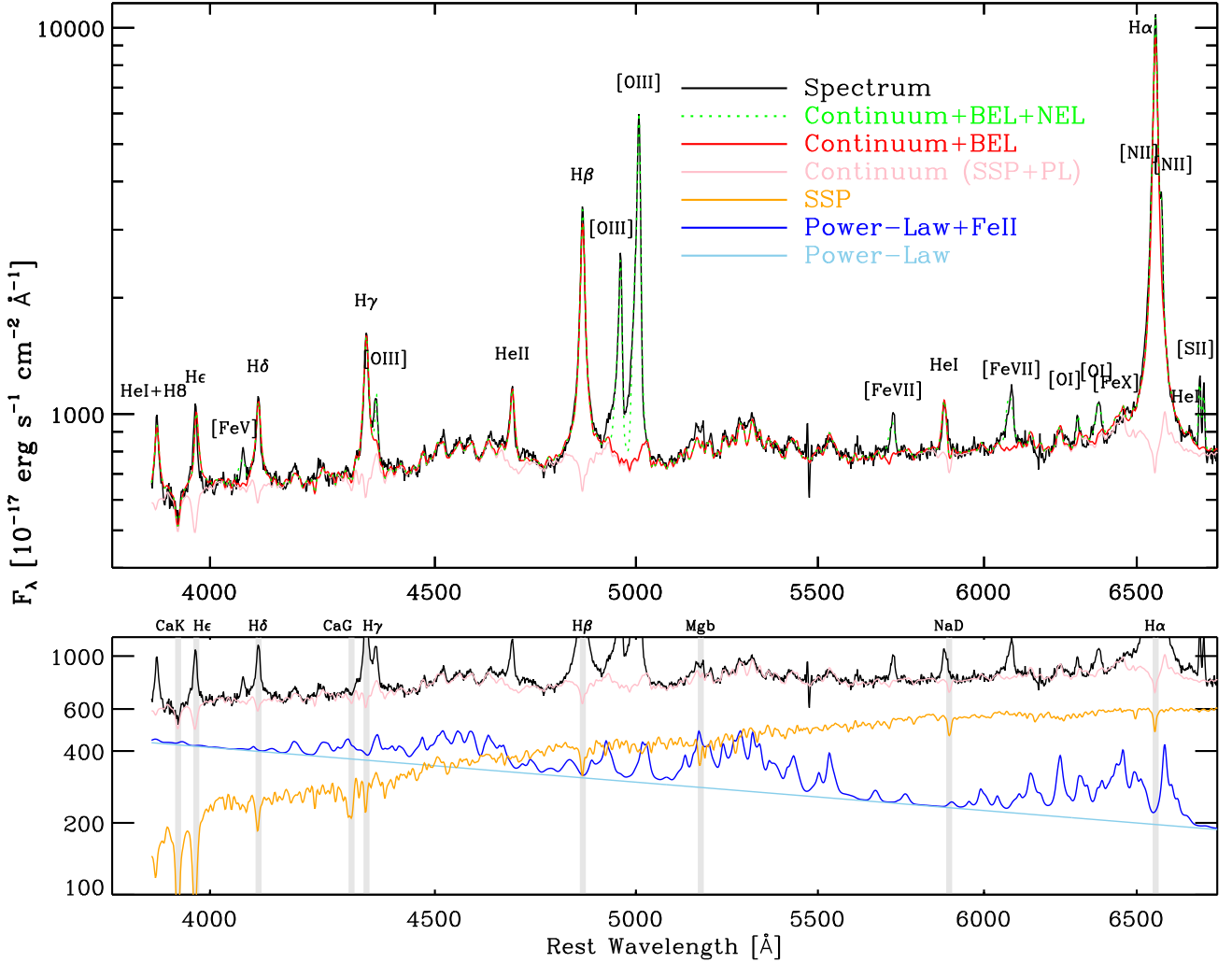}
\caption{Modeling of the 6dF low-resolution optical spectrum of Mrk 1239. The observed data are
shown in black lines. The best-fit models, i.e. continuum, continuum+BEL, and continuum+BEL+NEL are
overlaid in different colors respectively in the upper panel. In the lower panel, 
a zoom-in comparison between the data and modeled continuum (power law+\feii+SSP) components are shown,
with positions of prominent stellar absorption lines marked with vertical grey lines. }
\end{figure*}

The ESI optical spectra of the nucleus of Mrk 1239 is also fitted with a similar procedure. At a relatively small aperture, the underlying continuum of the ESI spectra appears to be much simpler, as stellar absorption lines are not obviously observed. We then model the continuum (power law + \feii multiplets) and emission lines iteratively as described above. Fitting results are shown in Figure 11. The $FWHM$ of the broad emission lines are found to be $822\pm 1$ \kms, similar to the results based on low-resolution data. On the other hand, some narrow emission lines appear very complicated and are not approximated well with the model. The $FWHM$ of the narrow peak is 350$\pm$1\kms,  with a broad base having $FWHM$ of 1467$\pm$2\kms and a shifting speed of $-447\pm2$\kms. The fitting result for the NELs are a bit different from the results based on low-resolution data. Since the narrow emission lines clearly show an irregular and blueshifted component, the two-Gaussian model is probably unable to fully recover the emission line profile. On the other hand, it is noticed that the broad emission lines of Mrk 1239 are reported to be wider in the literature, e.g. $FWHM_{\rm \pabeta} = 2200\pm133$ \kms (Marinello et al. 2016), $FWHM_{\rm \halpha} = 2375$ \kms (Rodríguez-Ardila et al. 2000). We think that the reasons lead to the different results can be complex. 1) The resolution and spectral quality can affect fitting results (the resolution of the IRTF spectra in Marinello et al. (2016) is R$\sim 830$, and the spectral quality is considerable lower than the TripleSpec data in this work.). 2) Modeling of emission lines are affected significantly by the placement of the underlying continuum, e.g. we find that line-fitting without considering \feii multiplets leads to a broader \halpha width (1070\kms for the 6dF data, and 940\kms for the ESI data). 3) The actual emission line profiles of Mrk 1239 can be complicated as compared with either the assumed Lorentzian or Gaussian profiles. Though line fitting can be tricky in these aspects, we adopt the widhth of BELs fitted in this paper because different spectra components (SSP, \feii multiplets) and multiple hydrogen and helium permitted lines are all taken into account in the fitting precedure, and the resulted width for the broad emission line component is similar based on both the low-resolution 6dF and the relatively high-resolution ESI data.


\begin{figure*}
\includegraphics[width=0.99\textwidth]{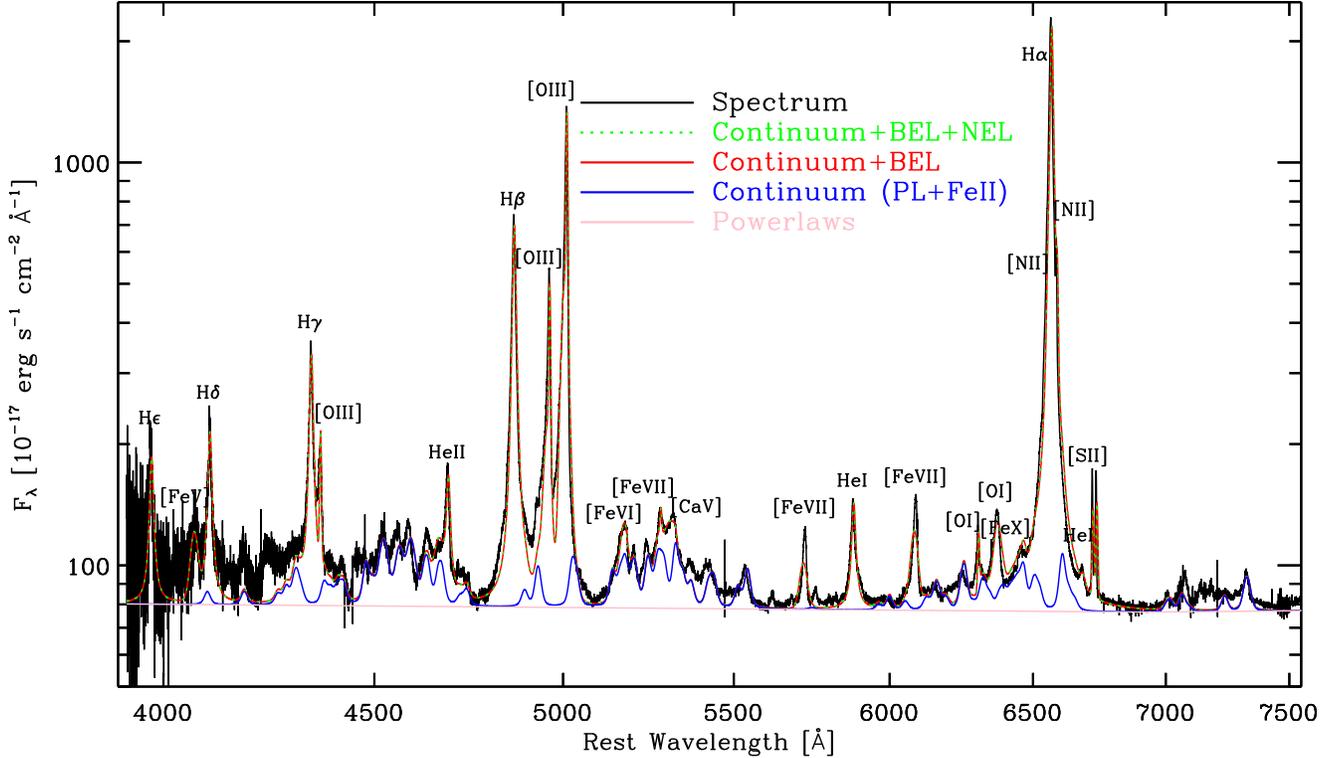}
\caption{Modeling of the Keck/ESI optical spectrum of Mrk 1239. The observed data are
shown in black lines. The best-fit model and its components, i.e. power-law continuum, continuum+BEL, and continuum+BEL+NEL are
overlaid in different colors respectively.}
\end{figure*}


The NIR spectrum of Mrk 1239 is relatively noisier and also simpler as compared with the optical band in both its continuum and emission lines. We first fit the data points in continuum windows with polynomials (green solid line in Figure 12(a)), and the polynomial model is then subtracted to obtain the emission line spectrum. As show in \S6, the narrow emission lines in Mrk 1239 are relatively weak in the NIR, thus a bit difficult to be taken into account. Luckily, Riffel et al. (2006) published NIR spectra of 14 Seyfert-2 galaxies, and the line spectra of these objects are dominated by narrow emission lines. After removal of the underlying polynomial continuum, emission line spectra of these targets in the NIR are then obtained (blue solid lines in Figure 12). We then model these emission lines with a single gaussian component and obtain the their flux ratios. The flux ratio templates are then applied to fit the isolated NELs in Mrk 1239 (grey shaded regions in Figure 12(a)). We then remove the NEL model (red solid line in Figure 12(a)) from the line spectra of Mrk 1239 (grey solid lines in Figure 12(b)) and fit permitted emission lines of hydrogen, \hei, \feii and \oi. We assign a Lorentzian profile for these permitted BELs and tie their centroid and width when fitting data points locally. Data points affected by \hei*$\lambda$10830~ BAL (pink-shaded region in Figure 12(b)) are masked during the fit. The best-fit result is shown with green solid lines in Figure 12(b). The width of the IR BELs are found to be $FWHM_{\rm IR~BEL} = 1090 \pm 80$ \kms, slightly wider than the optical BELs. Also, significant flux excess is found in the blue wing of \hei$*\lambda$10830, indicating the presence of potential outflowing emission line component.

\begin{figure*}
\includegraphics[width=0.99\textwidth]{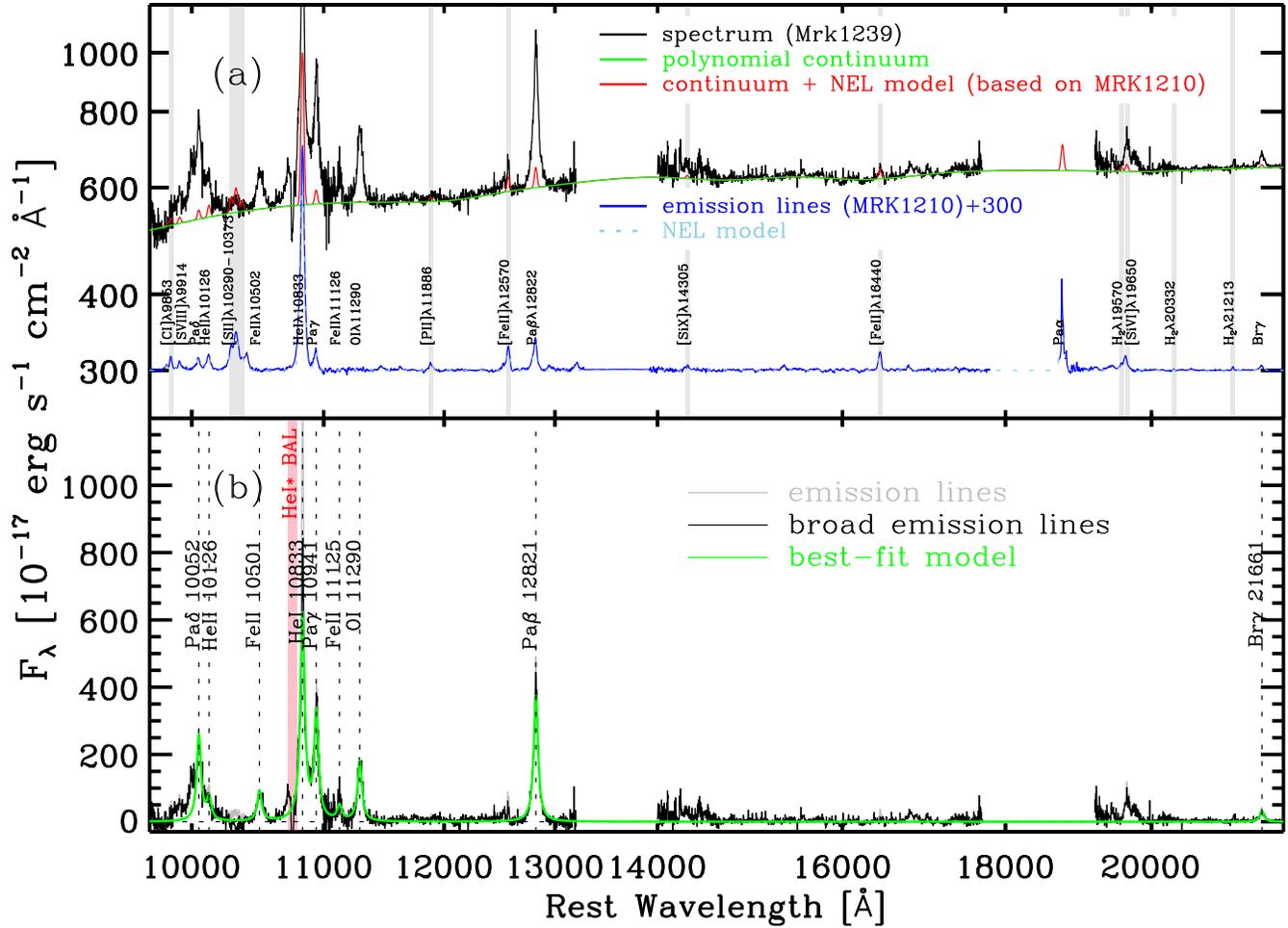}
\caption{Modeling of the NIR spectrum of Mrk 1239. In panel (a), the emission line spectra (blue solid lines) and the gaussian model (sky blue dashed lines) of a Seyfert-2 galaxy (Mrk 1210) is shown. After subtracting the underlying polynomial continuum, the isolated narrow emission lines (grey-shaded regions) of Mrk 1239 are fitted based on the line flux ratios of the template (best-fit result is in red solid lines). After removing continuum and NELs, the broad emission lines (black solid lines) are then fitted with a Lorentzian component, with the best-fit result shown in green solid lines. Note that data points affected by \hei*$\lambda$10830~ BAL (pink-shaded region) are masked during the fit.}
\end{figure*}

\section{A3. Photoionization modeling}

Based on the properties derived for the illuminant AGN (flux of the ionizing photon $Q_{\rm H} \approx 2.6 \times 10^{56}\rm\ s^{-1}$) and the outflowing gas ($N_{H}/\rm cm^{-2}$ = $22.0\sim22.9$, log $N_{\rm HeI (2^3S)}/\rm cm^{-2}$ = 13.5 $\pm$ 0.2 and log $N_{\rm HI n=2}/\rm cm^{-2} <$ 13.1), we carry out CLOUDY modelings to analyse the photoionization of the obscuring clouds. 
MF87 AGN SED is assumed as the incident radiation. Simulations with the ionization parameter $U$ ranging from 10$^{-4}$ to 1, and the hydrogen number density $n_{\rm H}$ ranging from 1 cm$^{-3}$ to 10$^{10}$ cm$^{-3}$, are carried out, because we only have column density measurements of hydrogen and helium, whose abundances are not affected by the metallicity of the cloud. 
Similar to the studies of quasar intrinsic absorbers, we fix the metallicity at solar abundance. Two sets of CLOUDY modellings with stopping criteria of log $N_{H}/\rm cm^{-2}$ = 22 and log $N_{H}/\rm cm^{-2}$ = 23.5 are carried out, and the inferred column density of the meta-stable He (n=$2^3S$) and H (n=2) are compared with measurements in Figure 13. We can see that despite the 1.5-dex difference of log $N_{H}/\rm cm^{-2}$, the results of the two sets of modellings for $N_{\rm HeI (2^3S)}$ indicate that the obscurer is at around 1 kilo-parsec distance to the illuminant AGN. 
However, the measured $N_{\rm HeI (2^3S)}$ is probably a lower limit owing to potential contamination from blueshifted \hei emissions ,
and the absorber may not be able to fully cover the AGN radiation in the NIR (e.g. Rodriguez-Ardila \& Mazzaly 2006). 
Nevertheless, the obscurer should be at distances lower than 1 kpc. On the other hand, the upper limit of log $N_{\rm HI n=2}/\rm cm^{-2} <$ 13.1 favors a distance of larger than $\sim$0.01 kpc from the central illuminant for relatively high hydrogen column density of around log $N_{H}/\rm cm^{-2}$ = 23.5. 
At a smaller hydrogen column density, though, the distance of the obscurer may be even smaller than 0.01 kpc if ionization parameters are high (log $U > -1.5$).
In general, CLOUDY simulations indicate that the distance of the obscurer to the illuminant AGN is around $R \sim [0.01,1]$ kpc.

\begin{figure*}
\includegraphics[width=\textwidth]{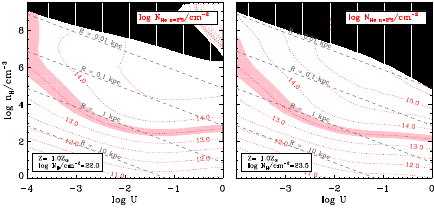}
\caption{
       CLOUDY simulations vs. the measured BAL column densities are shown. The hydrogen 
       column densities are log $N_{H}/\rm cm^{-2}$ = 22 and log $N_{H}/\rm cm^{-2}$ = 23.5 for 
       the left and the right panels respectively. Red dotted lines are contours of log $N_{\rm HeI (2^3S)}/\rm cm^{-2}$ 
       predicted by the model, with pink stripes indicating the measurements. 
       Regions in the parameter space rejected based on the upper limit measured for log $N_{\rm HI n=2}$ 
       is shaded with vertical lines. In general, CLOUDY simulations indicate a distance of 
       the obscurer to the illuminant AGN to be around $R \sim [0.01,1]$ kpc}
\end{figure*}

\end{document}